\documentstyle[prb,aps]{revtex}
\begin{document}
\draft
\title{Structure and metastability of superheated Al(111)}

\author{ G. Bilalbegovi\'c }

\address{Fakult\"at f\"ur Physik, Universit\"at Bielefeld,
D-33615 Bielefeld, Germany}

\date{\today}


\maketitle

\begin{abstract}

The high-temperature properties of the Al(111) surface are
studied by molecular-dynamics simulation.
This surface does not melt below the bulk melting point,
but can be superheated. Superheating of metal surfaces has 
been recently observed in several experiments. 
A molecular-dynamics study of the structural properties reveals 
how after going through the superheating regime melting occurs 
over the whole crystal in a narrow temperature range.
The temperature dependence of the surface stress, the mean-square 
vibrational amplitudes and the anomalous outward expansion of 
the distance between two top layers are calculated.
A transition from superheated to liquid state is analyzed using 
kinetic description for the formation of liquid nuclei by 
the Fokker-Planck equation and  conservation of heat at the liquid-solid 
interface.

\end{abstract}




\section{Introduction}

The question of how melting occurs, in spite of its long
history and importance, remains to a large extent unanswered.
Although various metastable phases of different materials
are present in nature, for some time it was
a common belief that metastability is
not possible at the solid-to-liquid transition. The argument was that
a free surface of any
material in contact with vapor acts as a nucleus of a liquid phase and 
suppresses superheating, i.e., nonmelting above the bulk melting 
temperature. Nevertheless, superheating was recently observed 
in several materials. For metal crystals it is especially difficult 
to achieve superheating. The viscosity of liquid metals is low and 
the liquid-solid interface rapidly propagates into the bulk. Several ways 
to suppress melting of metals were found. One way was to enclose 
the metal crystal into another metal with higher melting temperature. 
For example, superheating up to $62$ K was observed
for small Pb precipitates in Al \cite{Grabek}.
It was checked that the Pb particles of mean size $\sim 200$ {\AA}
remain superheated at $18$ K above the bulk melting point for more than
$21$ hours \cite{Grabek}. 
The Ag spheres were coated with Au and superheated by $25$ K 
for a period of about one minute \cite{Daeges}.
Such a behavior was modeled in molecular-dynamics (MD) simulation
using the Lennard-Jones potentials and different strength of interaction
\cite{Broughton}.

Techniques to superheat clean metal surfaces were also developed.
It is known that 
at the temperatures below the bulk melting point some metal surfaces 
exhibit surface melting.
Using medium--energy ion scattering it was found that the Pb(110) 
surface melts and that the thickness of liquid layer diverges
when the temperature approaches the bulk melting point
\cite{Amsterdam}.
On the contrary, Pb(111) does not melt and Pb(100) exhibits incomplete
melting characterized by the presence of liquid diffusion only in one
or two topmost layers \cite{Amsterdam}.
Similar behavior show low-index surfaces of other fcc metals, 
in particular surfaces of aluminum \cite{Vandergon}.
It was also found that some close-packed metal surfaces 
under special conditions may exhibit superheating. 
Herman and coworkers have found superheting by $120$ K for 
Pb(111) \cite{Herman}, $90$ K for Bi(0001) \cite{Murphy}, and even $15$ K
for incompletely melted Pb(100) \cite{Herman93}. Superheating 
of these surfaces was studied by time-resolved reflection high-energy 
electron diffraction. Melting was prevented by rapid heating
with a pulsed laser beam. Using laser pulses of the width $\sim 10^2$ ps
it was possible to bypass melting by inducing large heating and cooling 
rates of about $10^{11}$ K/s. 
By applying the same procedure it was not possible to superheat 
the Pb(110) surface \cite{Herman92,Herman94}. Superheating was also 
found in MD simulation of laser-pulse irradiation for Cu(111) 
\cite{Hakkinen}. For a development of new methods in  materials processing
by the laser beams it is important to understand the properties of
superheated surfaces.

Small crystallites bounded by nonmelting facets show superheating.
For example, nonequilibrium octahedral lead crystallites on graphite
(made up of the (111) facets and 
small round parts) exhibit superheating by several K 
and for several hours \cite{Metois}. 
Similar  superheating was also observed in MD high-temperature studies for
Au(111) \cite{Carnevali}, Al(111) \cite{Francesco}, 
Cu(111) \cite{Manninen}, and Pb(111) \cite{Goranka}.
DiTolla and coworkers have performed MD simulation for deposition
of a liquid aluminum cluster on the melting Al(110) and nonmelting Al(111)
surfaces \cite{Francesco}. 
They used these results and thermodynamic arguments to 
connect superheating with the wetting angle and the non-melting induced
faceting angle \cite{Goranka}.
The type of superheating related to fcc(111) 
is enabled by the exclusive presence of these non-melted 
surfaces on the specially prepared crystallites and on MD slabs. 
On equilibrium crystals the surface-melted facets (such as fcc(110)) 
are also present. They act as a nucleus of the liquid phase and prevent 
superheating of the whole crystal.

Till now the studies of superheating phenomena have been
mainly concentrated on the means to achieve the superheated state.
In this work MD simulation method was used to
study the properties of this metastable state, in particular 
the superheated Al(111) surface.
The analysis of the mean-square displacements, surface
relaxation, MD particle trajectories, and surface stress has been 
carried out.  The results of kinetic theory based on 
the Fokker-Planck equation
\cite{Landau} and  an analysis of the heat transport at the liquid-solid
interface \cite{Langer,Muller-Krumbhaar} 
were used to estimate the maximum superheating temperature for aluminum.
This analysis employed several quantities that were experimentally determined
and deduced from MD simulations. 
In the following the MD simulation method is described in Sec. II. 
Results of simulation and discussion are presented in Subsec. III A. 
Subsection III B deals with the kinetic analysis of transition 
from the superheated to liquid state. 
A summary and conclusions are given in Sec. IV.

\section{Molecular-dynamics computations}

The structure of the aluminum surface at high temperatures
was studied by MD simulation. The interatomic interactions    
were derived from a classical many-body 
potential \cite{Furio}. This form of potentials gives a proper
physical picture of metallic bonding \cite{Daw}.
The optimal set of parameters in the potential was found by 
the force-matching fitting to   
{\it ab initio} electronic structure calculations using a numerical
optimization procedure \cite{Furio}. 
The data used for fitting
were generated from different geometries, both at $T=0$ K
and for finite temperatures. As a result the potential is characterized
by  a good transferability.  
This potential was already used in MD simulations and it  
reproduced well  experimental results 
for bulk and surface properties of aluminum 
\cite{Francesco,Furio,Xqwang}.
The melting point  $T_m=939 \pm 5$ K is
in good agreement with the experimental value of 
$933.52$ K \cite{Web}. The calculated bulk melting
temperature was precisely determined by simulating coexisting
liquid and solid phases under constant energy \cite{Jim,Orio}.

The simulation of the high-temperature properties of the Al(111) surface
started at $T=0$ K.
The MD box of 1600 particles was used. These atoms were arranged in 
the usual single slab geometry with the thickness of $N_z =16 (111)$ layers 
(i.e., $\sim 35 \AA$), and a $10 \times 10$ square lattice in each layer.
As in other classical MD studies of surfaces, the slab was extended along
$x$ and $y$ by using periodic boundary conditions and no boundary 
conditions were used along $z$. 
The three bottom layers of the slab were kept fixed to simulate the bulk.
The lattice constant 
was changed with temperature according to the expansion coefficient found 
in MD simulation under zero pressure.
The time step of $2.64 \times 10^{-15}$ s was used.
Most runs were carried out at the constant temperature (i.e., in the 
canonical ensemble). The temperature was controlled by rescaling the particle 
velocities at each time step. 
At each temperature at least
$10^4$ time steps were performed to ensure 
thermal equilibration. In the superheating regime the length of 
the runs was much longer, i.e., up to $10^6$ time steps. 
In this regime simulations at fixed energy were also performed and
the same results as in the canonical ensemble were obtained.

The stress analysis often shows important changes of surface 
properties. The surface stress tensor $\sigma_{ij}$ is given by
\begin{equation}
\sigma_{ij}=\gamma \delta_{ij} + \frac{\partial\gamma}
{\partial\epsilon_{ij}},
\label{eq:1}
\end{equation}
where $\gamma$ is the surface free energy per unit area, 
$\delta_{ij}$ is 
the Kronecker symbol, $\epsilon_{ij}$ is the surface
strain tensor, and $i$, $j$ are directions in the surface plane
\cite{Cammarata}. 
The surface stress can be  obtained from the components of the
pressure tensor calculated during MD simulation \cite{Gilmer}.
In the computation of the stress MD boxes equilibrated for 
at least $10^5$ time steps were used. 

\section{Results and Discussion}

\subsection{Structure}

The studies of the various
structural properties of the Al(111) surface (such as 
the density, the static structure factor, the orientational order 
parameter, and MD particle trajectories) were done. 
The detailed analysis of these properties between
$0$ K and $1500$ K shows that at $\sim 1120$ K
melting starts on a typical MD simulation time scale.
The temperature of $1050$ K was selected for a detailed 
analysis of the superheating regime. At this temperature 
the surface stays superheated after $10^6$ time steps. 
The static structure factor, the density and order parameter plots
(not presented here), as well as the MD particle trajectories
in Fig.~\ref{fig1} show that 
the superheated Al(111) surface at $1050$ K and after $10^6$ time 
steps is crystalline and well ordered.
It was found that in the superheating regime single
adatoms often appear above the surface (see Fig.~\ref{fig1} (b)).
These adatoms have a very short life time ($\sim 10^{-12}$ s).
This is in contrast with the results
obtained using the same type of potentials for the high-temperature 
behavior of fcc(110) (surface melting) \cite{Simoneta}
and fcc(100) (incomplete melting) \cite{Erio}.
For these surfaces  the formation of adatoms was not observed. 
The Al(111) surface at T=$1120$ K stays superheated for a 
long time of $4\times 10^5$ MD steps. In the interval of $(40-45)\times
10^4$ time steps melting begins and develops. It was found that at this 
temperature melting proceeds with different speed in different MD runs.
For example, while in one run eleven layers were melted after 
$42\times 10^4$ MD steps, in another one after $45\times 10^4$ 
time steps only five top layers of MD box were melted.
Internal energy as a function of temperature is shown in Fig.~\ref{fig2}.
The jump region on the caloric curve between the
crystalline and liquid states is less abrupt than usually for 
melting transitions. The maximum superheating temperature estimated 
in this simulation is $180$ K. DiTolla and coworkers reported 
$\sim 150$ K for the same surface, using the same potential and similar 
simulation times \cite{Francesco}. 
This is a result of  different shape and size of MD boxes used 
in these simulations. A  dependence of the calculated bulk melting
temperature on the size, shape, and orientation of the samples is 
well known problem in MD simulation \cite{Jim,Orio}.

The surface stress was calculated and its temperature dependence is
presented in Fig.~\ref{fig3}. The (111) surface is isotropic, 
the $xx$ and $yy$ components of the stress tensor are approximately equal 
and therefore the stress is represented by their mean values.
The stress of $0.056$ eV {\AA}$^{-2}$ was found
for the relaxed surface at $T=0$ K.
This is $3.3$ times less than the average stress for Au(110) at 
the same temperature and for the same type of potential \cite {Wetting}.
The calculated stress for Al(111) at $T=0$ K is 
in excellent agreement with 
$0.059$ eV {\AA}$^{-2}$ obtained for the same surface in MD simulation 
using the Sutton-Chen potential \cite{Ruth}.
These MD results should be compared with two 
{\it ab initio} electronic structure calculations for Al(111). 
The value of $0.078$ eV {\AA}$^{-2}$ was found by Needs and Godfrey 
\cite{Needs} and $0.090$ eV {\AA}$^{-2}$ by Feibelman  \cite{Feibelman}.
Schmid and coworkers calculated surface stress for some low-index fcc 
metal surfaces using MD simulation and effective medium theory potentials 
\cite{Schmid}. They found that MD generally gives lower values  
compared to electronic structure calculations.
Figure 3 shows that surface stress for Al(111) is almost constant
between $T=0$ K and $T=900$ K. The stress is always tensile.
The components of any type of stress (or pressure) tensor calculated in 
MD simulation exhibit large fluctuations \cite{Rowlinson}. In Fig. 3 
the errors increase with the temperature and
the maximum of statistical uncertainty 
for the points is $20 \%$.
Same trend in the distribution of the values 
(i.e., that the stress remains almost 
constant over the wide temperature region and also that pronounced 
scatter of the data exists) was found in MD simulation  
for the (111) surface of the Lennard-Jones crystal \cite{Gilmer}.
It is well known that surface stress [see Eq. (1)] for a liquid
becomes equal to the surface free energy (i.e., ${\partial\gamma}/
{\partial\epsilon_{ij}}=0$). 
The value $0.05$ eV {\AA}$^{-2}$, obtained in this 
simulation for the liquid surface at $T=1500$ K, is equal to 
the surface free energy of liquid aluminum \cite{Amsterdam}.

The mean-square vibrational amplitudes  for the surface atoms 
are shown in Table \ref{table1}. The values of mean-square 
displacements are presented at two temperatures below 
the bulk melting point ($300$ K and $900$ K) and also
at $1050$ K, i.e., for a typical temperature 
in the superheating region.  
The last row of this table is obtained for the solid 
surface at the onset of melting at $T=1120$ K.
The values found for temperatures below the bulk melting point 
are in a good agreement with other MD simulation results 
\cite{Erio,Liqiu,Beaudet}. 
The mean-square vibrational amplitudes
at the end of the superheating region (not studied elsewhere)
are $\sim 10$ times larger than below the bulk melting temperature. 
The mean-square amplitudes of vibration are isotropic at 
all temperatures. 
Most MD simulations of the high-temperature properties of 
fcc metal surfaces
also show that the mean-square vibrational amplitudes are isotropic
\cite{Liqiu,Beaudet}.
In recent experimental studies \cite{Jiang} and MD simulation \cite{Erio}
of two fcc(100) surfaces it was found that
the mean-square amplitudes of vibration are anisotropic. 
Out-of-plane vibrational amplitudes were found to be
smaller than in-plane ones.
This reveals lateral disordering typical for incomplete melting. 
Isotropic mean-square vibrational amplitudes found here for Al(111)
show that after superheating regime melting proceeds in similar 
way along all directions.

Most low-index metal surfaces relax inwards at lower temperatures.
It is known that the Al(111) surface exhibits an anomaly in relaxation,
i.e., that at low temperatures
the distance between the two outmost layers expands in 
comparison with the bulk layers \cite{Adams}. The potential for aluminum
used in this paper gives good description of the  Al(111) relaxation,
i.e., it gives the experimental value for surface relaxation
$+ 0.9\%$ at $T=90$ K \cite{Furio,Adams}. In this work the temperature 
dependence of surface relaxation was studied and results are also 
presented in Table \ref{table1}. It was found that this
unusual outward expansion of the distance between the two top layers
increases with temperature. This increase is $\sim 1\%$ along 
the superheating temperature region. The maximum of surface relaxation 
is $+3.3\%$ at the onset of melting. 
 
\subsection{Metastability}

The fundamental questions of conditions and limits for the existence 
of any metastable state deserve further investigation. For metastable 
Al(111) it is important to consider a maximum of the superheating
temperature as a function of aluminum properties.
For fcc metal crystals the maximum of superheating should occur at the
close packed non-melted (111) surfaces. 
MD simulation presented above shows 
that  the mean-square vibrational amplitudes are isotropic 
at the transition from the superheated to liquid state
(see Table \ref{table1}). Moreover, drops form everywhere,
and the liquid front rapidly propagates into the bulk. 
In the following estimate 
of the maximum superheating temperature small anisotropy 
of the solid-liquid interface free energy  
is not taken into account. Therefore, the spherical
liquid nuclei are analyzed.

The kinetic theory of first order phase transitions 
provides a description of
transformation from a metastable to stable phase \cite{Landau}. 
This transition proceeds
via fluctuation induced formation of the nuclei of the stable phase. 
If the
radius $r$ of the nucleus is smaller than some critical value $r_c$ such
nucleus disappears, if it is bigger the nucleus grows. The radius of the
spherical liquid critical nucleus is 
\begin{equation}
r_c=\frac {2\alpha} {L} \frac{T_m} {T-T_m},
\label{eq:2}
\end{equation}
where $\alpha$ is the solid-liquid interface free energy  
and $L$ is the latent 
heat of melting \cite{Lan5}. It is possible to describe the growth of 
the nucleus by the kinetic Fokker-Planck equation for the distribution
function $f(r,t)$
\begin{equation}
\frac{\partial f} {\partial t}=-\frac {\partial w} {\partial r},
\label{eq:3}
\end{equation}
where $w$ is the flux density. 
The phase transition corresponds to the stationary 
solution of Eq. (3), where $w=const$. Such solution is \cite{Landau}
\begin{equation}
w=2\sqrt \frac {\alpha} {T} B(r_c) f_0(r_c).
\label{eq:4}
\end{equation}
In this equation $B$ is diffusion type coefficient and $f_0(r_c)$ is
\begin{equation}
f_0(r_c)=const \exp \left (\frac {-4\pi \alpha r_c^2} {3 T} \right).
\label{eq:5}
\end{equation}
Equation  (4) gives the rate for formation of a critical nucleus. 
The goal here is to calculate 
the maximum of superheating, i.e., the temperature at which melting 
occurs. 
Therefore, at this temperature $w\sim 1$. The coefficient $B(r_c)$ 
in Eq. (4),  is given by \cite{Landau}
\begin{equation}
B(r)=\frac {T} {8\pi\alpha(r-r_c)} \frac {dr} {dt},
\label{eq:6}
\end{equation}
and can be calculated [for $r\rightarrow r_c$ as in Eq. (4)] 
by considering a particular metastable state.

For a transition from the superheated to liquid state 
it is necessary to consider
the diffusive transport of heat at the liquid-solid interface 
\cite{Langer,Muller-Krumbhaar}.
It is convenient to define the dimensionless field
\begin{equation}
u=\frac{C_p} {L} (T_\infty - T),
\label{eq:7}
\end{equation}
where $T_\infty$ is the temperature of the solid infinitely far from the
growing nucleus and $C_p$ is the specific heat. Then the diffusion
equation is 
\begin{equation}
\frac {\partial u} {\partial t}= D \bigtriangledown ^2 u,
\label{eq:8}
\end{equation}
where D is the thermal diffusion constant at $T\sim T_m$.
One boundary condition is the heat conservation at the liquid-solid 
interface
\begin{equation}
v_n= -D \hat{\vec {n}} \cdot \bigtriangledown  u,
\label{eq:9}
\end{equation}
where $\hat{\vec {n}}$ is the unit normal directed outward from the 
nucleus and
$v_n$ is the normal growth velocity. There is also a requirement of local 
thermodynamic equilibrium that gives the dimensionless temperature $u_l$
at the interface
\begin{equation}
u_l= \bigtriangleup - d_0 {\cal K},
\label{eq:10}
\end{equation}
where 
\begin{equation}
\bigtriangleup = \frac {C_p} {L} (T_\infty - T_m)
\label{eq:11}
\end{equation}
is proportional to superheating. In the second term in Eq. (10) 
(i.e., in the Gibbs-Thomson correction for the melting temperature 
at a curved 
surface) $\cal K$ is the curvature and $d_0= \alpha C_p T_m L^{-2}$.
The solution for the growth rate of a spherical nucleus is 
\cite{Muller-Krumbhaar}
\begin{equation}
\left . \frac {dr} {dt} \right | _{r\rightarrow r_c} \sim  \frac {D} {r^2} 
(\bigtriangleup r_c - 2 d_0).
\label{eq:12}
\end{equation}
Therefore, using Eqs. (5), (6), (12), 
and $T_\infty = T$, condition $w\sim 1$ for Eq. (4) gives
\begin{equation}
a T^{1/2} (T-T_m)^3 \exp\{-[b/T(T-T_m)^2]\}=1,
\label{eq:13}
\end{equation}
where $a={DLC_p}/ {16\pi {\alpha}^{5/2} T_m^2}$ and
$b= {16 \pi {\alpha}^3 T_m^2}/ {3L^2}$.
Equation (13) was solved numerically using the experimental data for 
interface free energy $\alpha$ \cite{Amsterdam},
latent heat of melting $L$ \cite{Amsterdam}, 
bulk melting temperature $T_m$ \cite{Web}, 
specific heat $C_p$ \cite{Web}, 
and diffusion constant $D$ \cite{Ludwig}. 
The value $\sim 23$ K was obtained for the maximum
of superheating. This value is smaller than the maximum superheating 
temperature of $180$ K estimated in MD simulation for a model of the
Al(111) surface.  
Part of this disagreement is the result of the approximate kinetic
analysis and even possibly poor accuracy of some experimental data for Al.
There is also a possibility that the value for the maximum of superheating
found in MD simulations is in part the result of a limited time evolution.
Although long runs of $10^6$ time steps were performed in the
superheating regime, much longer (nowadays not feasible) simulation
times may give the maximum of the superheating temperature in accordance 
with the kinetic result.
Using the kinetic analysis presented above, the same literature sources
for $\alpha$, $L$, $T_m$, $C_p$, and diffusion constant $D$ from Ref. 
\cite{Nachtrieb}, the value of $\sim 47$ K was obtained for 
the maximum superheating temperature of Pb.
For aluminum a similar Fokker-Planck analysis was done using quantities 
deduced from the simulations.
As in Ref. \cite{Ruth}, it was found that the bulk energy is slightly 
nonlinear function of temperature and that therefore it can be fitted by 
the second-order polynomial. Using this procedure \cite{Ruth}, the value
of $C_p = 1120$ J/(K kg) was obtained, whereas the experimental value for
the specific heat is $902$ J/(K kg) [23]. For diffusion constant the
value of $D=0.3 \times 10^{-5}$ $cm^2/s$ was found (see also \cite{Franz}). 
This is much lower than the experimental value $3 \times 10^{-5}$ 
$cm^2/s$ \cite{Ludwig}. When in Eq. (13) the quantities obtained in 
MD simulations for $C_p$, $D$, $L=0.105$ eV/atom \cite{Furio},  
$T_m=939$ K \cite{Furio}, and $\alpha = 10$ $meV \AA^{-2}$ 
\cite{Francesco,Franz} were used, then $48$ K was calculated for the maximum 
superheating temperature of aluminum.
It is important to point out that smaller values for the maximum 
superheating temperature were found in the experiments on the crystallites 
\cite{Grabek,Daeges,Metois}, whereas larger superheating was induced
by a laser beam \cite{Herman,Murphy,Herman93,Herman94}.

\section{Summary and Conclusions}

The properties of a surface in the superheated state
were studied using MD simulation and a reliable
many-body interatomic potential. 
Superheated Al(111) was used as a model of superheated surfaces, such as
ones obtained by a pulsed laser beam
\cite{Herman,Murphy,Herman93,Herman94}, or on the crystallites 
\cite{Metois}. A detailed analysis of 
the Al(111) surface from room temperature to $1500$ K (i.e., well above
the bulk melting point) was carried out. The results for lower 
temperatures are in a good agreement with available experimental 
observations and other MD simulations and electronic structure 
calculations for metal surfaces.
It is possible to superheat the sample by $\sim 180$ K
for typical longest simulation times used in classical MD ($> 2.5 $ ns). 
In the superheating regime the Al(111) surface is remarkably well ordered,
although single adatoms sometimes appear.
The sample melts over the narrow temperature interval. 
Anomalous outward expansion between two top layers  increases slowly 
with the temperature: from $+0.9\%$ at $T=0$, 
up to $+3.3\%$ at the end of the superheating region. The mean-square 
vibrational amplitudes are isotropic for all temperatures and $\sim 10$ 
times larger at the end of
the superheating region than below the bulk melting point.
It was shown that kinetic theory based on the Fokker-Planck equation and
analysis of heat conservation at the liquid-solid interface for aluminum 
gives the maximum superheating temperature of $23$ K when experimentally
determined parameters where used. The maximum superheating temperature of
$48$ K was obtained when parameters deduced from MD simulations 
where applied in the Fokker-Planck analysis. This analysis of kinetics 
and process of disordering observed in MD simulation shows
that superheated Al(111) and bulk Al below the surface are 
an example of metastability at the 
solid-to-liquid transition. The kinetics of this transition
is the same as in other better known examples of metastability 
\cite{Landau}.
Superheated surfaces of other metals should exhibit similar behavior.

\acknowledgments

I would like to thank  F. Ercolessi, B. Gumhalter, and E. Tosatti
for discussions.

\clearpage

\begin{table}
\caption{
Mean-square vibrational amplitudes for the Al(111) surface
(in units of {\AA}$^2$):
$u^2_x$ 
(along the $[1\bar{1}0]$ direction),  $u^2_y$ (along
the $[11\bar{2}]$ direction) and $u^2_z$ (along the vertical axis).
The surface relaxation $d_{12}$ between the two
top layers is also shown (in {\%}).}
\label{table1} 
\begin{tabular}{l l l l l} 
Temperature  & $u_x^2$ 
 & $u_y^2$ & $u_z^2$  & $d_{12}$ \\ 
\hline
$300$ K & $0.014$ & $0.022$ & $0.018$ & $+ 1.1$ \\ 

$900$ K & $0.074$ & $0.064$ & $0.075$ & $+ 2.2$\\ 

$1050$ K & $0.118$ & $0.119$ & $0.101$ & $+ 2.4$ \\ 

$1120$ K  & $0.839$ & $0.708$ & $0.731$ & $+ 3.3$ \\ 
\end{tabular}
\end{table}


\begin{figure}
\caption{
Particle trajectories showing the superheated surface after $10^6$ 
MD steps
of time evolution at $1050$ K  (i.e., $110$ K above the
bulk melting temperature):
(a) top view, (b) side view with an adatom above the surface.
Trajectory plots refer to a time span of $\sim 3$ ps
and include only moving atoms.}
\label{fig1}
\end{figure}

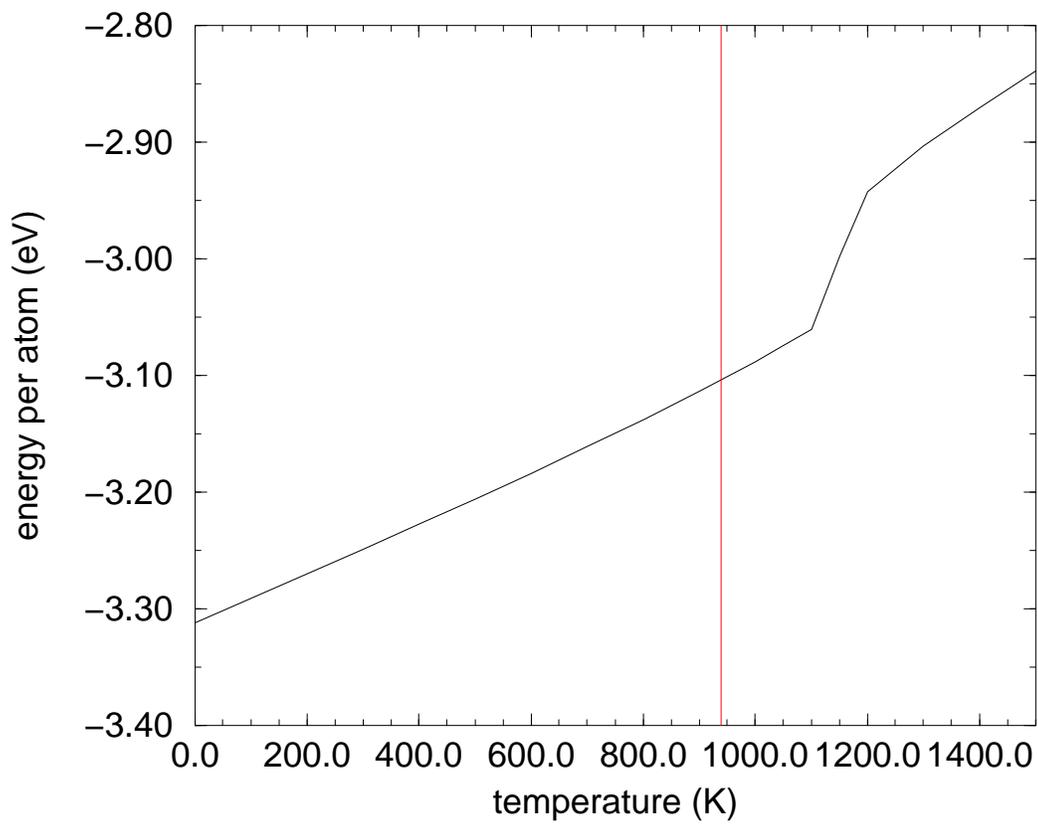
\begin{figure}
\caption{
The caloric curve. The vertical line represents 
the calculated bulk melting temperature.}
\label{fig2}
\end{figure}

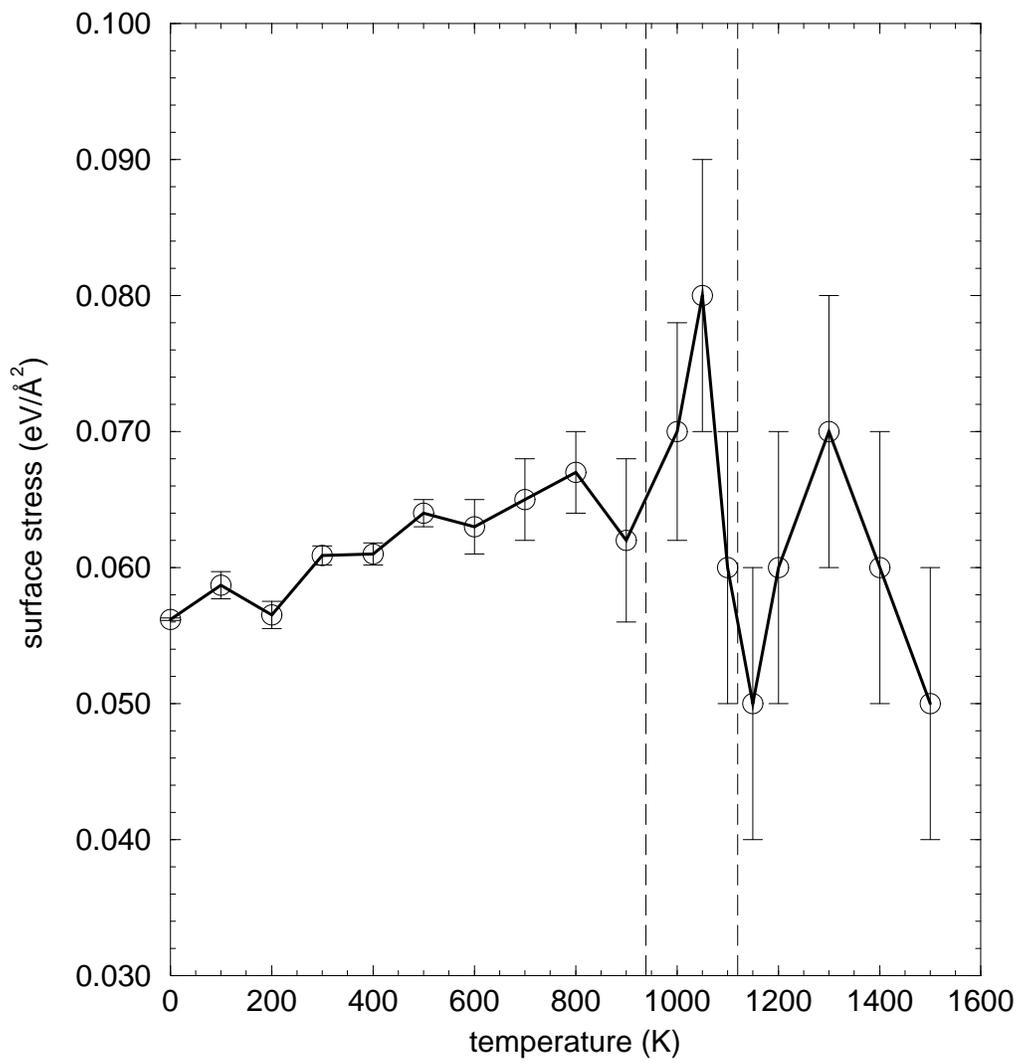
\begin{figure}
\caption{
Surface stress as a function of temperature. The dashed vertical lines
enclose the superheating region.}
\label{fig3}
\end{figure}

\end{document}